\documentclass[conference]{IEEEtran}
\IEEEoverridecommandlockouts
\usepackage{cite}
\usepackage{amsmath,amssymb,amsfonts}
\usepackage{algorithmic}
\usepackage{glossaries}
\usepackage{graphicx}
\usepackage{textcomp}
\usepackage{xcolor}
\usepackage{lipsum}
\usepackage{gensymb}
\usepackage{glossaries}
\usepackage{multirow}
\usepackage{makecell}
\usepackage{color, colortbl}
\definecolor{mixed}{gray}{0.9}
\definecolor{synth}{gray}{0.8}
\usepackage{multicol}
\usepackage[tableposition=top]{caption}
\usepackage{subcaption}
\usepackage{siunitx}
\usepackage{balance}
\usepackage{colortbl}
\usepackage{pgfplots}
\usepackage{pgfplotstable}
\usepgfplotslibrary{fillbetween}
\usepgfplotslibrary{external}
\usepackage{longtable}
\usepackage{xspace}
\usepackage{arydshln}
\usepackage{booktabs}

\usepackage{soul}
\usepackage{url}

\usepackage[font=scriptsize]{subcaption}
\usepackage[font=footnotesize]{caption}

\usepackage{amsmath}

\usepackage[outdir=./]{epstopdf}
\usepackage{tikz}
\pgfplotsset{compat=newest}
\pgfplotsset{plot coordinates/math parser=false}
\newlength\fheight
\newlength\fwidth
\usetikzlibrary{plotmarks,shapes,patterns,decorations.pathreplacing,backgrounds,calc,arrows,arrows.meta,spy,matrix} 
\usepackage{tikzscale}
\usepackage[utf8]{inputenc}

\IEEEoverridecommandlockouts
\newcommand\copyrightnotice{%
\begin{tikzpicture}[remember picture,overlay]
\node[anchor=south,yshift=10pt] at (current page.south) {\fbox{\parbox{\dimexpr\textwidth-\fboxsep-\fboxrule\relax}{
\footnotesize \textcopyright 2025 IEEE. Personal use of this material is permitted.
Permission from IEEE must be obtained for all other uses, in any current or future media,
including reprinting/republishing this material for advertising or promotional purposes,
creating new collective works, for resale or redistribution to servers or lists,
or reuse of any copyrighted component of this work in other works.}}};
\end{tikzpicture}
}

\usepackage[capitalise]{cleveref}
\crefname{section}{Sec.}{Secs.}
\crefname{figure}{Fig.}{Figs.}

\usepackage{eqparbox}
\def\BibTeX{{\rm B\kern-.05em{\sc i\kern-.025em b}\kern-.08em
    T\kern-.1667em\lower.7ex\hbox{E}\kern-.125emX}}

\newacronym{3gpp}{3GPP}{3rd Generation Partnership Project}
\newacronym{5g}{5G}{5th Generation}
\newacronym{5gc}{5GC}{5G Core}
\newacronym{6g}{6G}{6th Generation}
\newacronym{adc}{ADC}{Analog to Digital Converter}
\newacronym{afbw}{AFBW}{Average Fading Bandwidth}
\newacronym{aimd}{AIMD}{Additive Increase Multiplicative Decrease}
\newacronym{am}{AM}{Acknowledged Mode}
\newacronym{amc}{AMC}{Adaptive Modulation and Coding}
\newacronym{aoa}{AoA}{Angle of Arrival}
\newacronym{aod}{AoD}{Angle of Departure}
\newacronym{ap}{AP}{Access Point}
\newacronym{app}{APP}{Application Layer}
\newacronym{aqm}{AQM}{Active Queue Management}
\newacronym{awgn}{AGWN}{Additive White Gaussian Noise}
\newacronym{balia}{BALIA}{Balanced Link Adaptation}
\newacronym{bdp}{BDP}{Bandwidth-Delay Product}
\newacronym{ber}{BER}{Bit Error Rate}
\newacronym{bler}{BLER}{Block Error Rate}
\newacronym{bf}{BF}{Beamforming}
\newacronym{cad}{CAD}{Computer-Aided Design}
\newacronym{cbr}{CBR}{Constant Bit Rate}
\newacronym{cc}{CC}{Congestion Control}
\newacronym{cdf}{CDF}{Cumulative Distribution Function}
\newacronym{ci}{CI}{Confidence Interval}
\newacronym{cir}{CIR}{Channel Impulse Response}
\newacronym{cn}{CN}{Core Network}
\newacronym{cp}{CP}{Control Plane}
\newacronym{cqi}{CQI}{Channel Quality Information}
\newacronym{crs}{CRS}{Cell Reference Signal}
\newacronym{csirs}{CSI-RS}{Channel State Information - Reference Signal}
\newacronym{d2d}{D2D}{Device-to-Device}
\newacronym{dc}{DC}{Dual Connectivity}
\newacronym{dce}{DCE}{Direct Code Execution}
\newacronym{dci}{DCI}{Downlink Control Information}
\newacronym{dl}{DL}{Downlink}
\newacronym{dmr}{DMR}{Deadline Miss Ratio}
\newacronym{dmrs}{DMRS}{DeModulation Reference Signal}
\newacronym{dray}{D-Ray}{Deterministic Ray}
\newacronym{e2e}{E2E}{End-to-End}
\newacronym{ecn}{ECN}{Explicit Congestion Notification}
\newacronym{ecdf}{ECDF}{Empirical Cumulative Distribution Function}
\newacronym{edf}{EDF}{Earliest Deadline First}
\newacronym{em}{EM}{electromagnetic}
\newacronym{enb}{eNB}{evolved Node Base}
\newacronym{endc}{EN-DC}{E-UTRAN-\gls{nr} \gls{dc}}
\newacronym{epc}{EPC}{Evolved Packet Core}
\newacronym{es}{ES}{Edge Server}
\newacronym{eess}{EESS}{Earth Exploration-Satellite Service}
\newacronym{fdd}{FDD}{Frequency Division Duplexing}
\newacronym{fdma}{FDMA}{Frequency Division Multiple Access}
\newacronym{fray}{F-Ray}{Flashing Ray}
\newacronym{fs}{FS}{Fast Switching}
\newacronym{fss}{FSS}{Fixed Satellite Service}
\newacronym{ftp}{FTP}{File Transfer Protocol}
\newacronym{gmm}{GMM}{Gaussian Mixture Model}
\newacronym{gnb}{gNB}{Next Generation Node B}
\newacronym{gr}{GR}{Ground Reflection}
\newacronym{harq}{HARQ}{Hybrid Automatic Repeat reQuest}
\newacronym{hetnet}{HetNet}{Heterogeneous Network}
\newacronym{hh}{HH}{Hard Handover}
\newacronym{hol}{HOL}{Head-of-Line}
\newacronym{hpbw}{HPBW}{Half Power Beamwidth}
\newacronym{hqf}{HQF}{Highest-quality-first}
\newacronym{ia}{IA}{Initial Access}
\newacronym{iab}{IAB}{Integrated Access and Backhaul}
\newacronym{ieee}{IEEE}{Institute of Electrical and Electronics Engineers}
\newacronym{imt}{IMT}{International Mobile Telecommunication}
\newacronym{inr}{INR}{Interference to Noise Ratio}
\newacronym{iot}{IoT}{Internet of Things}
\newacronym{itu}{ITU}{International Telecommunication Union}
\newacronym{ked}{KED}{Knife-Edge Diffraction}
\newacronym{kpi}{KPI}{Key Performance Indicator}
\newacronym{ks}{KS}{Kolmogorov–Smirnov}
\newacronym{lcf}{LCF}{Level Crossing Frequency}
\newacronym{lcr}{LCR}{Level Crossing Rate}
\newacronym{leo}{LEO}{Low-Earth Orbit}
\newacronym{los}{LoS}{Line-of-Sight}
\newacronym{lte}{LTE}{Long Term Evolution}
\newacronym{m2m}{M2M}{Machine to Machine}
\newacronym{mac}{MAC}{Medium Access Control}
\newacronym{mc}{MC}{Multi-Connectivity}
\newacronym{mcl}{MCL}{Minimum Coupling Loss}
\newacronym{mcs}{MCS}{Modulation and Coding Scheme}
\newacronym{mec}{MEC}{Mobile Edge Cloud}
\newacronym{mi}{MI}{Mutual Information}
\newacronym{mib}{MIB}{Master Information Block}
\newacronym{mimo}{MIMO}{Multiple Input, Multiple Output}
\newacronym{mlr}{MLR}{Maximum-local-rate}
\newacronym{mls}{MLS}{Microwave Limb Sounder}
\newacronym[plural=\gls{mme}s,firstplural=Mobility Management Entities (MMEs)]{mme}{MME}{Mobility Management Entity}
\newacronym{mmwave}{mmWave}{millimeter wave}
\newacronym{moi}{MoI}{Method of Images}
\newacronym{mpc}{MPC}{Multi Path Component}
\newacronym{mptcp}{MPTCP}{Multipath TCP}
\newacronym{mr}{MR}{Maximum Rate}
\newacronym{mrdc}{MR-DC}{Multi \gls{rat} \gls{dc}}
\newacronym{mss}{MSS}{Maximum Segment Size}
\newacronym{mtd}{MTD}{Machine-Type Device}
\newacronym{mtu}{MTU}{Maximum Transmission Unit}
\newacronym{nfv}{NFV}{Network Function Virtualization}
\newacronym{nist}{NIST}{National Institute of Standards and Technology}
\newacronym{nlos}{NLoS}{Non-Line-of-Sight}
\newacronym{nr}{NR}{New Radio}
\newacronym{nrmse}{NRMSE}{Normalized Root Mean Square Error}
\newacronym{ns3}{ns-3}{Network Simulator 3}
\newacronym{nsa}{NSA}{Non Stand Alone}
\newacronym{ntn}{NTN}{Non Terrestrial Network}
\newacronym{o2i}{O2I}{Outdoor-to-Indoor}
\newacronym{ofdm}{OFDM}{Orthogonal Frequency Division Multiplexing}
\newacronym{osm}{OSM}{OpenStreetMap}
\newacronym{pa}{PA}{Position-aware}
\newacronym{pbch}{PBCH}{Physical Broadcast Channel}
\newacronym{pdcch}{PDCCH}{Physical Downlonk Control Channel}
\newacronym{pdcp}{PDCP}{Packet Data Convergence Protocol}
\newacronym{pdf}{PDF}{Probability Density Function}
\newacronym{pdsch}{PDSCH}{Physical Downlink Shared Channel}
\newacronym{pdu}{PDU}{Packet Data Unit}
\newacronym{per}{PER}{Packet Error Rate}
\newacronym{pf}{PF}{Proportional Fair}
\newacronym{pgw}{PGW}{Packet Gateway}
\newacronym{phy}{PHY}{Physical}
\newacronym{pl}{PL}{Path Loss}
\newacronym{ppp}{PPP}{Poisson Point Process}
\newacronym{prb}{PRB}{Physical Resource Block}
\newacronym{pss}{PSS}{Primary Synchronization Signal}
\newacronym{pucch}{PUCCH}{Physical Uplink Control Channel}
\newacronym{pusch}{PUSCH}{Physical Uplink Shared Channel}
\newacronym{qd}{QD}{Quasi Deterministic}
\newacronym{rach}{RACH}{Random Access Channel}
\newacronym{ran}{RAN}{Radio Access Network}
\newacronym[firstplural=Radio Access Technologies (RATs)]{rat}{RAT}{Radio Access Technology}
\newacronym{red}{RED}{Random Early Detection}
\newacronym{rf}{RF}{Radio Frequency}
\newacronym{rfi}{RFI}{Radio Frequency Interference}
\newacronym{rlc}{RLC}{Radio Link Control}
\newacronym{rlf}{RLF}{Radio Link Failure}
\newacronym{rr}{RR}{Round Robin}
\newacronym{rray}{R-Ray}{Random Ray}
\newacronym{rrc}{RRC}{Radio Resource Control}
\newacronym{rrm}{RRM}{Radio Resource Management}
\newacronym{rs}{RS}{Remote Sensing}
\newacronym{rsrp}{RSRP}{Reference Signal Received Power}
\newacronym{rsrq}{RSRQ}{Reference Signal Received Quality}
\newacronym{rss}{RSS}{Received Signal Strength}
\newacronym{rssi}{RSSI}{Received Signal Strength Indicator}
\newacronym{rt}{RT}{Ray Tracer}
\newacronym{rtt}{RTT}{Round Trip Time}
\newacronym{rw}{RW}{Receive Window}
\newacronym{rx}{RX}{Receiver}
\newacronym{sa}{SA}{standalone}
\newacronym{sack}{SACK}{Selective Acknowledgment}
\newacronym{sap}{SAP}{Service Access Point}
\newacronym{sbr}{SBR}{Shooting-and-Bouncing Rays}
\newacronym{sch}{SCH}{Secondary Cell Handover}
\newacronym{scm}{SCM}{Spatial Channel Model}
\newacronym{scoot}{SCOOT}{Split Cycle Offset Optimization Technique}
\newacronym{sdma}{SDMA}{Spatial Division Multiple Access}
\newacronym{sf}{SF}{Shadow Fading}
\newacronym{si}{SI}{Study Item}
\newacronym{sib}{SIB}{Secondary Information Block}
\newacronym{sinr}{SINR}{Signal-to-Interference-plus-Noise Ratio}
\newacronym{sir}{SIR}{Signal-to-Interference Ratio}
\newacronym{sm}{SM}{Saturation Mode}
\newacronym{snr}{SNR}{Signal-to-Noise Ratio}
\newacronym{son}{SON}{Self-Organizing Network}
\newacronym{srs}{SRS}{Sounding Reference Signal}
\newacronym{ss}{SS}{Synchronization Signal}
\newacronym{sss}{SSS}{Secondary Synchronization Signal}
\newacronym{sta}{STA}{Station}
\newacronym{subthz}{sub-THz}{sub-TeraHertz}
\newacronym{svd}{SVD}{Singular Value Decomposition}
\newacronym{tb}{TB}{Transport Block}
\newacronym{tcp}{TCP}{Transmission Control Protocol}
\newacronym{udp}{UDP}{User Datagram Protocol}
\newacronym{tdd}{TDD}{Time Division Duplexing}
\newacronym{tdma}{TDMA}{Time Division Multiple Access}
\newacronym{te}{TE}{Transverse Electric}
\newacronym{tfl}{TfL}{Transport for London}
\newacronym{tgad}{TGad}{Task Group ad}
\newacronym{tgay}{TGay}{Task Group ay}
\newacronym{tm}{TM}{Transverse Magnetic}
\newacronym{trp}{TRP}{Transmitter Receiver Pair}
\newacronym{tti}{TTI}{Transmission Time Interval}
\newacronym{ttt}{TTT}{Time-to-Trigger}
\newacronym{tx}{TX}{Transmitter}
\newacronym{ue}{UE}{User Equipment}
\newacronym{ul}{UL}{Uplink}
\newacronym{um}{UM}{Unacknowledged Mode}
\newacronym{uma}{UMa}{Urban Macro}
\newacronym{uml}{UML}{Unified Modeling Language}
\newacronym{utc}{UTC}{Urban Traffic Control}
\newacronym{vm}{VM}{Virtual Machine}
\newacronym{wbf}{WBF}{Wired Bias Function}
\newacronym{wf}{WF}{Wired-first}
\newacronym{wifi}{Wi-Fi}{Wireless Fidelity}
\newacronym{wigig}{WiGig}{Wireless Gigabit}
\newacronym{wlan}{WLAN}{Wireless Local Area Network}
\newacronym{xpr}{XPR}{Cross Polarization Ratio}
\newacronym{sthz}{Sub-THz}{sub-terahertz}
\newacronym{thz}{THz}{terahertz}
\newacronym{fr}{FR}{Frequency Range}
\newacronym{cbrs}{CBRS}{Citizen Broadband Radio Service}
\tikzstyle{startstop} = [rectangle, rounded corners, minimum width=2cm, minimum height=0.5cm,text centered, draw=black]
\tikzstyle{io} = [trapezium, trapezium left angle=70, trapezium right angle=110, minimum width=3cm, minimum height=1cm, text centered, draw=black]
\tikzstyle{process} = [rectangle, minimum width=2cm, minimum height=0.5cm, text centered, draw=black, alignb=center]
\tikzstyle{decision} = [ellipse, minimum width=2cm, minimum height=1cm, text centered, draw=black]
\tikzstyle{arrow} = [thick,<->,>=stealth]
\tikzstyle{line} = [thick,>=stealth]
\tikzstyle{darrow} = [thick,<->,>=stealth,dashed]
\tikzstyle{sarrow} = [thick,->,>=stealth]
\tikzstyle{larrow} = [line width=0.1mm,dashdotted,->,>=stealth]

\makeatletter
\def\grd@save@target#1{%
  \def\grd@target{#1}}
\def\grd@save@start#1{%
  \def\grd@start{#1}}
\tikzset{
  grid with coordinates/.style={
    to path={%
      \pgfextra{%
        \edef\grd@@target{(\tikztotarget)}%
        \tikz@scan@one@point\grd@save@target\grd@@target\relax
        \edef\grd@@start{(\tikztostart)}%
        \tikz@scan@one@point\grd@save@start\grd@@start\relax
        \draw[minor help lines] (\tikztostart) grid (\tikztotarget);
        \draw[major help lines] (\tikztostart) grid (\tikztotarget);
        \grd@start
        \pgfmathsetmacro{\grd@xa}{\the\pgf@x/1cm}
        \pgfmathsetmacro{\grd@ya}{\the\pgf@y/1cm}
        \grd@target
        \pgfmathsetmacro{\grd@xb}{\the\pgf@x/1cm}
        \pgfmathsetmacro{\grd@yb}{\the\pgf@y/1cm}
        \pgfmathsetmacro{\grd@xc}{\grd@xa + \pgfkeysvalueof{/tikz/grid with coordinates/major step x}}
        \pgfmathsetmacro{\grd@yc}{\grd@ya + \pgfkeysvalueof{/tikz/grid with coordinates/major step y}}
        \foreach \x in {\grd@xa,\grd@xc,...,\grd@xb}
        \node[anchor=north] at (\x,\grd@ya) {\pgfmathprintnumber{\x}};
        \foreach \y in {\grd@ya,\grd@yc,...,\grd@yb}
        \node[anchor=east] at (\grd@xa,\y) {\pgfmathprintnumber{\y}};
      }
    }
  },
  minor help lines/.style={
    help lines,
    gray,
    line cap =round,
    xstep=\pgfkeysvalueof{/tikz/grid with coordinates/minor step x},
    ystep=\pgfkeysvalueof{/tikz/grid with coordinates/minor step y}
  },
  major help lines/.style={
    help lines,
    line cap =round,
    line width=\pgfkeysvalueof{/tikz/grid with coordinates/major line width},
    xstep=\pgfkeysvalueof{/tikz/grid with coordinates/major step x},
    ystep=\pgfkeysvalueof{/tikz/grid with coordinates/major step y}
  },
  grid with coordinates/.cd,
  minor step x/.initial=.5,
  minor step y/.initial=.2,
  major step x/.initial=1,
  major step y/.initial=1,
  major line width/.initial=1pt,
}
\makeatother

\begin{document}


\title{Spectrum Sharing Across Terrestrial and Non-Terrestrial Services in the FR3 Upper Midband}

\author{
\IEEEauthorblockN{Paolo Testolina, Ergest Beshaj, Michele Polese, Tommaso Melodia}
\IEEEauthorblockA{Institute for the Wireless Internet of Things, Northeastern University, Boston MA}
\IEEEauthorblockA{\{p.testolina, beshaj.e, m.polese, melodia\}@northeastern.edu}
\thanks{This work was partially supported by the U.S. NSF under award CNS-2332721 and by OUSD(R\&E) through Army Research Laboratory Cooperative Agreement Number W911NF-24-2-0065. The views and conclusions contained in this document are those of the authors and should not be interpreted as representing the official policies, either expressed or implied, of the Army Research Laboratory or the U.S. Government. The U.S. Government is authorized to reproduce and distribute reprints for Government purposes notwithstanding any copyright notation herein.}
}

\maketitle

\copyrightnotice

\begin{abstract}

The frequency bands between 7 and 24 GHz, also known as upper midband or \gls{fr} 3, are being considered as an enabler of \gls{6g} mobile networks. This portion of the spectrum exhibits different propagation characteristics compared to frequencies above 24 GHz, while also offering the potential to provide larger bandwidth allocations for mobile systems than those available in the sub-6 GHz range. \gls{6g} technology and spectrum policy, however, will need to guarantee coexistence with the incumbents that already use these frequency bands, which include a variety of services, from radiolocation to satellite-based communications, remote sensing, and radio astronomy. In this paper, we consider the challenge of coexistence between \gls{6g} terrestrial systems and satellite incumbents in different portions of the FR3 bands. Using a large-scale 3D model of a terrestrial deployment in the city of Boston and an open-source ray tracing solution, we evaluate the level of \gls{rfi} that tens of terrestrial \glspl{gnb} generate toward satellites at different elevation angles. Our model, based on realistic obstruction, clutter, diffraction, and reflections, shows that sidelobes and \gls{nlos} paths can significantly contribute to \gls{rfi}. Besides directionality, the spatial distribution of \glspl{gnb} also plays a key role in defining the \gls{rfi} levels, suggesting that a careful design and operation of terrestrial deployments can create coexistence opportunities.


\end{abstract}

\begin{tikzpicture}[remember picture,overlay]
\node[anchor=north,yshift=-10pt] at (current page.north) {\parbox{\dimexpr\textwidth-\fboxsep-\fboxrule\relax}{
\centering\footnotesize This paper was presented at the 2025 IEEE International Symposium on Dynamic Spectrum Access Networks (DySPAN). \textcopyright 2025 IEEE.\\
Please cite it as: P. Testolina, E. Beshaj, M. Polese and T. Melodia, ``Spectrum Sharing Across Terrestrial and Non-Terrestrial Services in the FR3 Upper Midband,'' 2025 IEEE International Symposium on Dynamic Spectrum Access Networks (DySPAN), London, United Kingdom, 2025, pp. 1-9, doi: 10.1109/DySPAN64764.2025.11115947.}};
\end{tikzpicture}

\section{Introduction}

\glsresetall


With the \gls{6g} standardization process underway within the \gls{3gpp}, industry, academia, and policymakers are looking into what spectrum should be used to support next-generation wireless systems~\cite{matinmikko2020spectrum}. While the lower \gls{mmwave} spectrum, or \gls{fr} 2, was seen as a promising enabler of ultra-high data rates in \gls{5g} systems, the challenges associated with supporting efficient and cost-effective end-to-end \gls{5g} deployments in these bands have limited their application in a limited number of dense, urban markets, leaving the promise unfulfilled~\cite{narayanan2022comparative}. At the same time, providing mobile service over large bandwidths (e.g., 400 MHz carriers, compared to 100 MHz or smaller blocks available at FR1) remains an enticing proposition, as it enables high area capacity as well as support for precise sensing and positioning applications being considered for \gls{6g}~\cite{giordani2020toward,uusitalo20216g}.

For this reason, the upper midband, or FR3, between 7 and 24 GHz, is considered as a candidate for \gls{6g} systems as it comes with opportunities to allocate larger chunks of bandwidth to mobile services, and has more favorable propagation characteristics compared to FR2~\cite{kang2024cellular,kang2024terrestrial,cui20236g}. At the same time, this portion of the spectrum comes with its own set of policy and engineering challenges. While being discussed as a single frequency range, different subportions have different design requirements in terms of, for example, antenna and beamforming support. Further, the upper midband spectrum is already allocated and in use for multiple, heterogeneous services, spanning from fixed satellite uplink and downlink connectivity to sensing and radiolocation solutions.

Therefore, any solution for allocating spectrum to \gls{6g} mobile services will be more effective if it incorporates a spectrum-sharing approach. Sharing allows new services to coexist with incumbents safely and does not lead to complex and extensive spectrum refarming processes, which usually have a significant economic and societal impact~\cite{furuya2015radio}. 
When considering sharing with legacy infrastructure, sharing mechanisms and options are more limited compared to a clean slate design, in particular if the incumbent infrastructure does not support coordination mechanisms~\cite{bhattarai2016overview}. For example, \gls{cbrs} in the sub-6 GHz spectrum evacuates the communication channels when an incumbent is detected, implementing a time-based sharing without the need for two-way coordination.

Another option is spatial sharing, where the different services can coexist thanks to intrinsic limits in the propagation domain. Specifically, when considering communications in higher frequency bands, e.g., FR3 and FR2, the directionality required to close the terrestrial link for mobile services can, in some way, facilitate spatial sharing with satellite or non-terrestrial incumbents. FR3 hosts several different kinds of non-terrestrial incumbents, including communication links (which receive interference from terrestrial networks in their uplink receivers), sensing, and radio astronomy. Each of the services come with different needs and requirements in terms of sensitivity and protection from \gls{rfi}.

In this paper, we provide insights on whether spatial sharing is possible within terrestrial and non-terrestrial (specifically, \gls{leo}) networks at FR3. Compared to recent literature in this area~\cite{kang2024terrestrial,kang2024cellular}, we consider two key elements of next-generation terrestrial cellular systems. The first is the high density of the deployments, with a large number of terrestrial \glspl{gnb} generating potential \gls{rfi} to satellite incumbents. The second is directionality, i.e., we model beamforming with \gls{mimo} arrays, and its interplay with reflections originating from the ground and obstacle, to analyze how this affects the average and worst-case \gls{rfi}. We specifically focus on simulating the \gls{rfi} between the terrestrial system and the in-orbit incumbent of a non-terrestrial network, differently from~\cite{niloy2024ascent,niloy2023interference,liu2023research}.

The contributions of this paper are as follows. \textbf{(i)} We answer the question \emph{what are the network conditions and configurations that can lead to coexistence opportunities for terrestrial and non-terrestrial networks?} \textbf{(ii)} We do so by extending an open-source framework that combines large-scale ray tracing and 3D modeling~\cite{testolina2024bostontwin} to the non-terrestrial domain, modeling satellites and their orbits.
\textbf{(iii)} We leverage a 3D model based on real-world maps and building elevation to simulate large-scale \gls{rfi} in an urban scenario. This is usually associated with high deployment density but also comes with significant building obstructions to the \gls{los} between \glspl{gnb} and satellites, which facilitates the coexistence of terrestrial and non-terrestrial systems.
\textbf{(iv)} For the same reason, we show that \gls{los} channel models commonly employed for ground-to-satellite sharing studies and \gls{rfi} analysis might not capture the rich scattering conditions of the urban environment, where the multipath components play a key role.

Our results show that \textbf{(a)} the sidelobes of directional beams can decrease their effectiveness in suppressing the \gls{los} interference to the incumbent, if not carefully designed. \textbf{(b)} The \gls{nlos} components of the channel can deliver a non-negligible amount of interference to the satellite, even when the direct path is successfully attenuated, particularly when considering them in combination with directional communications systems. \textbf{(c)} The spatial distribution of \glspl{gnb} plays a pivotal role in the resulting \gls{rfi}, pointing to promising spatial-sharing schemes based on strategic placement or activation of the \glspl{gnb}. \textbf{(d)} Characterizing the aggregated interference from multiple sources is not trivial and is not always captured by the analysis of a single interferer.




\section{Related Works}
\label{sec:literature}

In the last decade, \gls{rfi} evaluation has been a key focus of research, due to the adoption of new frequency bands by commercial stakeholders. Several publications have provided insights~\cite{marcus2014harmful,chamberlain2024facilitating} to develop solutions for coexistence of current incumbent services and new cellular networks.

Many studies on \gls{rfi} follow the \gls{mcl} approach \cite{park2019modeling, polese2021coexistence, 9771541,kang2024cellular}, which analyzes the \gls{rfi} generated by a single interferer affecting a single victim. 
Kang et al.~\cite{kang2024cellular} focus on the evaluation of the \gls{rfi} to satellites in the FR3 bands, but consider a single link. 
Other recent papers focus on different frequency bands, e.g., above 100 GHz in~\cite{polese2021coexistence,xing2021terahertz}. 
Polese \textit{et al.} \cite{polese2021coexistence} review existing regulations for the spectrum above 100 GHz and highlight the importance of developing sharing strategies to optimize the spectrum usage in these frequency bands. The paper studies \gls{rfi} under a number of different scenarios aligned with ITU standards, such as addressing single backhaul terrestrial links.
%
Similarly, \cite{xing2021terahertz} considers an urban scenario, using a rooftop-mounted receiver as a surrogate for a satellite. The results indicate that in a single-link urban setting, the interference safety threshold at 140 GHz is maintained if the transmitter's beam stays under \ang{15} above the horizon. Our prior work~\cite{testolina2024modeling} extends the above-100-GHz analysis to multiple ground transmitters, showing the importance of considering ground reflections---amplified by the main lobe---in the analysis.

Other approaches aim to use stochastic tools, ray tracing, and propagation models to estimate the \gls{rfi}. For example, \cite{lim} analyzes aggregated interference in the $47.2-50.2$ GHz band and highlights the importance of considering the density of terrestrial transmitters, path loss, and stopband attenuation.
%
Similarly, \cite{ayoubi2023imt} introduces stochastic models that account for \gls{los}, reflections, diffraction, building clutter and foliage to characterize \gls{rfi} in urban scenarios, in the sub-6 GHz band. 
%
%
\cite{9556607} estimates the interference from terrestrial Fixed Service stations into aeronautical SATCOM links operating in the $17.7–19.7$ GHz band. The paper emphasizes that larger SATCOM antennas and higher flight altitudes can mitigate interference effectively. Similarly, \cite{winter2019statistics} examines the aggregated \gls{rfi} from a fixed-service network to an aircraft operating at 18~GHz.
Also at FR3,~\cite{kang2024terrestrial} investigates terrestrial-to-satellite spectrum sharing, considering the interference caused from cellular downlink transmission to the satellite uplink in the 12 GHz band and proposes a spatial nulling method to reduce \gls{rfi}. While~\cite{kang2024terrestrial} is based on a ray tracing tool, the evaluation does not consider an urban scenario, as we do in this paper, which may lead to different reflection and obstruction patterns.

\begin{figure*}[t]
    \centering
    \setlength\fheight{\columnwidth}
    \setlength\fwidth{2\columnwidth}
    \input{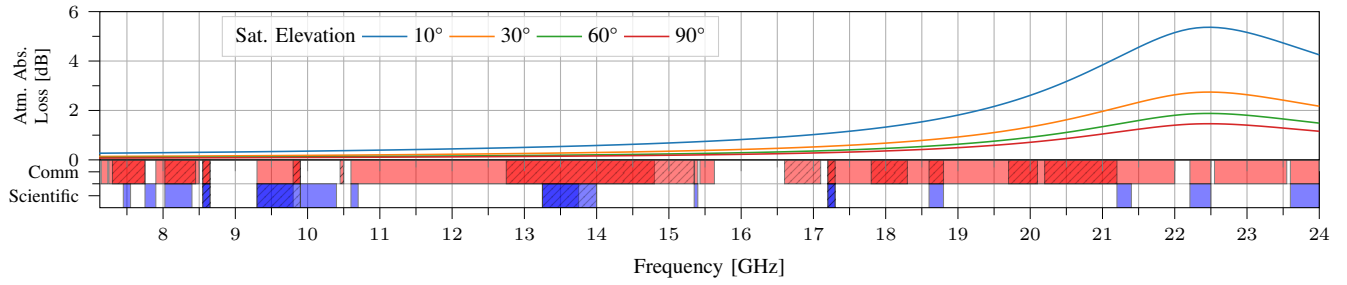}
    \caption{Overview of the spectrum allocations (bottom) and atmospheric absorption loss (top) across the upper-midband.}
    \label{fig:allocations}
\end{figure*}

Other papers that focus on \gls{rfi} in the upper midband include~\cite{niloy2023interference,niloy2024ascent,liu2023research,guidolin2015study,su2014coexistence}, which however consider a ground station as the target of the mobile network \gls{rfi}.
Niloy \textit{et al.} ~\cite{niloy2023interference} develop a simulation-based framework that incorporates realistic deployment scenarios, beamforming, directionality, and propagation models at 12 GHz, and highlight the importance of exclusion zones around satellite receivers. The authors of ~\cite{niloy2024ascent} dynamically adjust such zones with a closed-loop feedback system.
Similarly,~\cite{liu2023research} focuses on adjacent frequency interference and mitigation methods and emphasizes the need for isolation distances. 
The simulations in~\cite{guidolin2015study} and the findings in~\cite{su2014coexistence} evaluate the interference aggregated by \glspl{gnb} of nearby \glspl{imt} networks at 18 GHz and 3.4–3.6~GHz, targeting a \gls{fss} terrestrial station. Both studies conclude that coexistence between these systems is feasible at higher frequencies, provided that specific conditions for base station deployment and configuration are met. 
Authors in~\cite{gasiewski2002impacts} analyze the interference caused by automotive radars operating in the 22–27 GHz frequency range on satellite radiometers used for water vapor observations, showing that, under realistic scenarios, vehicle densities are likely to exceed \gls{rfi} thresholds.
%
Differently from these paper, we consider interference across different kinds of transmitters (terrestrial \glspl{gnb}) and receivers (satellites). 


Finally, \cite{zhong2020feasibility} studies interference from \gls{imt} networks to satellite relays in the 25.25–27.5~GHz bands, using \gls{los} propagation and the ITU channel model to aggregate interference over large areas. Similarly, \cite{cho2018spectral,cho2019modeling,cho2020coexistence} analyze the coexistence of terrestrial networks and \gls{eess} systems at \gls{mmwave} frequencies, accounting for \gls{rfi} from both the satellite’s 3-dB footprint and its broader coverage area. Ground nodes are modeled as geographically clustered \gls{5g} systems, with \glspl{gnb} and \gls{ue} antennas randomly placed and oriented. Aggregate \gls{rfi} is calculated by summing contributions from terrestrial links and clusters, incorporating factors such as free-space path loss, terrain elevation, and atmospheric attenuation. However, the simplified modeling of building blockage, attenuation, and beamforming may not fully capture real-world topologies, which are instead considered in this paper.

\section{System Model}
\label{sec:systemmodel}

Throughout this paper, we consider a dense terrestrial network deployment as the source of \gls{rfi}, and a non-terrestrial satellite-based system, which is subject to \gls{rfi}. 

\noindent\textbf{Terrestrial Network.} The terrestrial network represents a next-generation cellular deployment in an urban scenario, e.g., the city of Boston. Each \gls{gnb} transmits a directional signal to a local \gls{ue} within its coverage area, which is defined as an annulus with radiuses 5 and 400 m (corresponding to a pathloss of at most 130 dB, in the worst case).
The \glspl{ue} positions are determined as follows: 
\begin{enumerate}
    \item for each \gls{gnb}, a coverage map is computed;
    \item from each coverage map, the first 100 locations where the received power is the strongest are selected;
    \item at each iteration, a \gls{ue} is placed at a random  location among the pre-computed list of each \gls{gnb};
    \item the \gls{gnb} beam is steered toward to selected \gls{ue}.
\end{enumerate}
With this procedure, we ensure a realistic \gls{gnb}-\gls{ue} association and replicate realistic beamforming and beam steering dynamics.

The \glspl{gnb} are equipped with a single uniform planar array of $N_{tx}$ elements, where
\begin{equation}
    N_{tx} = \max{N^2} \text{ s.t. } \left(N\frac{\lambda}{2}\right)^2 \le A_{max},
\end{equation}
where $N$ represents the number of antenna elements along one dimension of the array, $\lambda$ is the wavelength corresponding to the carrier frequency, and $A_{max}$ is the array area, that we set as $40 \times 40$ mm$^2$.




\noindent\textbf{Non-Terrestrial Satellite System.} We model a satellite receiver transiting over the area of the terrestrial deployment, at different elevation angles, and with a beam that covers the whole area of the terrestrial deployment. 

International and national regulations reserve several bands for satellite-based services in the $7-24$~GHz range, as shown in Fig.~\ref{fig:allocations} (bottom).
The services can be classified as related to scientific, commercial, or national security missions. 
%
The first consists of satellites equipped with scientific instruments, e.g., the \glspl{eess}, that observe natural and anthropogenic phenomena for scientific purposes. The other services provide communications capabilities, which can be used to relay commercial, defense, or scientific data across the globe and to/from orbit and space missions, as well as radiolocation and radionavigation solutions.

A key difference between the scientific and the other classes of satellites is their sensitivity to \gls{rfi}. Scientific missions often aim at measuring signals with a very low \gls{snr}, and thus require extremely sensitive receivers, with a low tolerance for \gls{rfi}, especially in passive sensing contexts.
Conversely, communication systems or radiolocation and navigation generally have a higher tolerance to interference sources.
In this work, we propose an analysis that is incumbent-agnostic, i.e., we consider the signal at the input of the incumbent antenna.
In this way, we derive general results that can be easily extended to incumbent-specific considerations.


\section{Large-Scale \gls{rfi} Evaluation}
\label{sec:bostontwin}


To efficiently evaluate \gls{rfi} at large scale, we leverage BostonTwin~\cite{testolina2024bostontwin,testolina2024modeling,testolina2024sharing}, a digital twin of the city of Boston that combines (i) high-accuracy 3D models of the structures in the city; (ii) real-world locations of the \glspl{gnb}; and (iii) an optimized ray tracing toolchain. With this, it allows studying \gls{rfi} and other performance metrics over a large geographical area which includes realistic obstructions and deployment characteristics.
Large-scale digital twins are becoming the backbone of holistic network optimization, e.g., optimal \gls{gnb} deployment, coverage analysis, beam alignment, and interference management.
However, they are fundamental when considering any link between a ground and a satellite node, \emph{as the footprint of an antenna on board a satellite covers a large geographical area, even if highly directional.}

The BostonTwin encompasses a 360~km$^2$ area, which includes more than 1000 real \gls{gnb} locations and about 163000 structures, for a total of more than 11 million mesh triangles.


The BostonTwin is fully compatible with the Sionna ray-tracer~\cite{hoydis2023sionna}, a GPU-based, open-source ray-tracer developed by NVIDIA.
Reflection of arbitrary order, diffraction of the first order, and scattering on the considered meshes are simulated with a \gls{sbr} algorithm with a number of rays set by the user.
Compared to the (\textit{exhaustive}) \gls{moi}, which checks the impinging of the rays on every mesh of the scene by building a complete visibility tree~\cite{lecci2021accuracy}, \gls{sbr} launches a set of rays in different directions, traces the intersection of each ray to the triangles, up to the maximum reflection order and to the receiver.
Thus, the complexity of the latter is independent of the number of triangles, i.e., to the size and detail of the scene, but rather by the number of launched rays.
This is key to enabling ray-tracing in large-scale scenarios, reducing memory requirements and enabling detailed propagation environments.
However, when using the \gls{sbr}, the appropriate number of rays must be chosen, to guarantee that all the significant paths are explored.
Furthermore, the receiver is generally represented by a sphere rather than a point, to increase the probability of intersection and avoid numerical errors~\cite{egea2021opal,molina2024wireless}.
Both the receiver sphere radius and ray count must be chosen based on scene characteristics, such as scattering density, as no standard tuning method exists. To address this, Sionna refines \gls{sbr} results by applying \gls{moi} only to rays that hit the receiver.

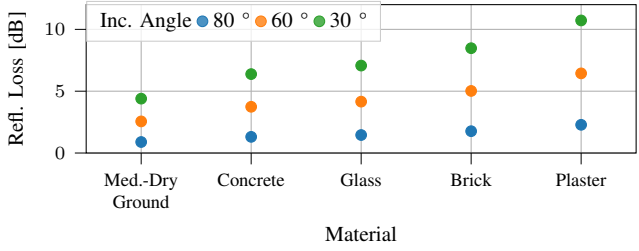
\begin{figure}[t]
    \centering
    \setlength\fheight{.4\columnwidth}
    \setlength\fwidth{\columnwidth}
    \begin{tikzpicture}
\pgfplotsset{every tick label/.append style={font=\scriptsize}}

\definecolor{crimson2143940}{RGB}{214,39,40}
\definecolor{darkgray176}{RGB}{176,176,176}
\definecolor{darkorange25512714}{RGB}{255,127,14}
\definecolor{forestgreen4416044}{RGB}{44,160,44}
\definecolor{lightgray204}{RGB}{204,204,204}
\definecolor{mediumpurple148103189}{RGB}{148,103,189}
\definecolor{sienna1408675}{RGB}{140,86,75}
\definecolor{steelblue31119180}{RGB}{31,119,180}

\begin{axis}[
 width=\fwidth,
  height=\fheight,
  at={(0\fwidth,0\fheight)},
  anchor=north west,
legend cell align={left},
legend style={
  fill opacity=0.8,
  draw opacity=1,
  text opacity=1,
  at={(0.0,0.75)},
  anchor=south west,
  draw=lightgray204,
  font=\footnotesize
},
legend columns=5,
tick align=outside,
tick pos=left,
x grid style={darkgray176},
xlabel={Material},
xmajorgrids,
xlabel style={font=\footnotesize},
ylabel style={font=\footnotesize},
xmin=0, xmax=10,
xtick style={color=black},
xticklabel style={align=center},
xtick={1,3,5,7,9},
xticklabels={{Med.-Dry\\ Ground}, Concrete, Glass, Brick, Plaster},
y grid style={darkgray176},
ymajorgrids,
ymin=0, ymax=12,
ytick style={color=black},
ylabel={Refl. Loss [dB]},
name=allocations,
]
\addlegendimage{empty legend}
\addlegendentry{Inc. Angle}

\addplot [draw=steelblue31119180, only marks, fill=steelblue31119180, mark=*]
table{%
x  y
1 0.8904859127136439
3 1.30752517184926
5 1.4552887986622238
7 1.764955161292563
9 2.2818037962077553
};
\addlegendentry{80~\degree}

\addplot [draw=darkorange25512714, only marks, fill=darkorange25512714, mark=*]
table{%
x  y
1 2.5562428041940333
3 3.739489677046265
5 4.155854851703507
7 5.020573678099202
9 6.441037266385431
};
\addlegendentry{60~\degree}

\addplot [draw=forestgreen4416044, only marks, fill=forestgreen4416044, mark=*]
table{%
x  y
1 4.397543327359136
3 6.382594634138632
5 7.0713463871592355
7 8.477064545173004
9 10.721367183968322
};
\addlegendentry{30~\degree}


\end{axis}

\end{tikzpicture}
    \setlength\belowcaptionskip{-10pt}
    \caption{The reflection loss has a negligible dependency on frequency across the upper midband.}
    \label{fig:refl_loss}
\end{figure}

\subsection{Ray-based Ground-to-Satellite \gls{rfi} Channel Model}


As part of this work, we have extended the integration between BostonTwin and the Sionna \gls{rt} to derive a ground-to-satellite channel model that takes into account different propagation phenomena. In this extension, these include the atmospheric absorption loss, which is dependent on the elevation angle and the frequency of the ground-to-satellite link.
The extension is open-source and available within the BostonTwin repository~\cite{testolina2024bostontwin}. 

In geometrical optics \glspl{rt}, the propagation of the \gls{em} field in the environment is represented by rays, which discretize the (continuous) field into the multiple geometric paths that connect the transmitter to the receiver.
Each ray carries part of the transmitted signal's energy and is characterized by its amplitude $a\in\mathcal{C}$, which depends on its interaction with the environment, \gls{aod} ($\phi^D,\theta^D$), \gls{aoa} ($\phi^A,\theta^A$), and delay $\tau$.
\emph{We include the direct ray and the multipath components generated by multiple reflections and by the first-order diffraction on the buildings of the considered area.} This leads to a highly accurate channel model, capable of capturing geometry-specific elements of the deployment, reflection across obstacles and the ground. This can be used to map specific transmitter on the ground to their impact on the \gls{rfi}, as we show in Sec.~\ref{sec:results}. 

\paragraph{Free-space Propagation}
As the signal propagates for a distance $l$, part of its energy is dissipated according to the free space loss coefficient $a_{fsp}(l,f)\in\mathcal{C}$.

\paragraph{Atmospheric Absorption}
The gases present in the Earth's atmosphere can interact with the \gls{em} field if its wavelength has a comparable dimension to the gas molecules.
In that case, the latter can resonate and dissipate part of the signal energy as thermal energy.
The \gls{itu} recommendations~\cite{itu-p-676} provide a standard way to model this interaction, and the pycraf package offers an excellent Python implementation of the standard\footnote{\url{https://github.com/bwinkel/pycraf}}~\cite{pycraf}.
The atmospheric absorption coefficient $a_{atm}$ depends on a number of atmospheric parameters, including humidity, temperature, and the resolution to model the atmospheric layers.
For this work, we use the mid-latitude summer height profile~\cite{itu-p-835}.
To include the atmospheric absorption in our model, we consider the elevation angle and altitude of the satellite and apply the corresponding atmospheric attenuation coefficient to the rays.
The atmospheric absorption loss for the upper midband and different elevation angles of a satellite orbiting the Earth at $400$~km is reported in \cref{fig:allocations}.

\paragraph{Reflection Loss}
When an \gls{em} wave impinges on a surface, part of the energy is absorbed by the surface itself.
The energy loss is modeled through the reflection attenuation coefficient $a_{refl}\in\mathcal{C}$, which depends on the material properties~\cite{itu-r-2040} and on the incident angle.

Note that in general the material's \gls{em} properties, and thus the reflection coefficient, depend on the frequency of the considered signal.
However, for the frequencies under consideration, the dependency on the frequency is almost negligible.
Figure~\ref{fig:refl_loss} reports the reflection coefficient for typical materials in the upper mid-band~\cite{itu-r-2040}.

\paragraph{Diffraction Loss}
Finally, the \gls{em} field can be diffracted by sharp edges present in the environment, creating diffraction patterns behind the obstacle.
Diffraction is generally modeled in geometrical-optics \glspl{rt} by bending the rays around the edges.
The diffraction coefficient is $a_{diff}$.

The \gls{em} field $E$ at any point $P$ in space is thus
\begin{equation}
    E(P) = \sum\limits_{r=0}^{R-1} a_r(P) E_0,
\end{equation}
where $R$ is the number of paths connecting the source to $P$, $E_0$ is the complex amplitude of the electric field at the source, and $a_r(P)$ is the total attenuation coefficient for ray $r$. The latter is computed as (dropping the dependency on $P$ for readability)
\begin{align}
    a_r = a_{r,fsp} a_{r,atm} a_{r,refl} a_{r, diff} \in \mathcal{C},
    \label{eq:total_abs}
\end{align}
with
\begin{equation}
    a_{r,refl} = \begin{cases}
    a_{refl} & \text{if $r$ is reflected}\\
    0 & \text{otherwise}
    \end{cases},
    \label{eq:refl_loss}
\end{equation}
and analogously for $a_{r,diff}$.

For this work, we consider a time-invariant channel, and we assume the symbol duration is long enough for the received multipath components to be aggregated in a single tap, thus neglecting the time dimension from the equations and from the analysis.

\begin{figure}[t]
    \centering
    \setlength{\fwidth}{\columnwidth}
    \setlength{\fheight}{.5\columnwidth}
    \input{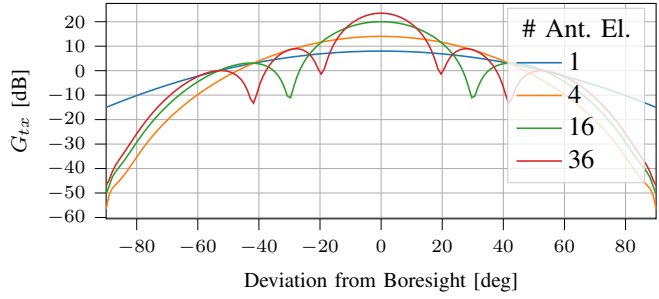}
    \caption{Transmitter gain for different deviation angles (x-axis) and array size. As the number of elements increases, the pattern becomes more directional, although sidelobes emerge.}
    \label{fig:pattern}
\end{figure}

\begin{figure*}[t]
\centering
    \begin{subfigure}[t]{\columnwidth}
        \centering
        \includegraphics[width=.7\columnwidth,clip, trim=0 0 0 10cm]{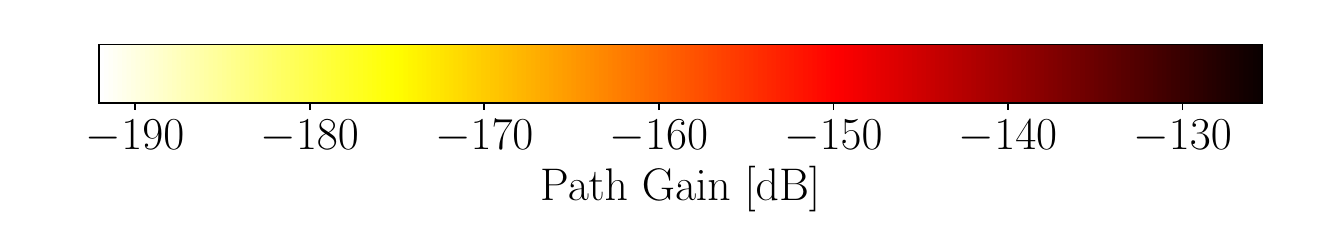}
    \end{subfigure}\\
    \begin{subfigure}[t]{0.65\columnwidth}
        \centering
        \includegraphics[width=\columnwidth,clip, trim=0 3.4cm 0 0cm]{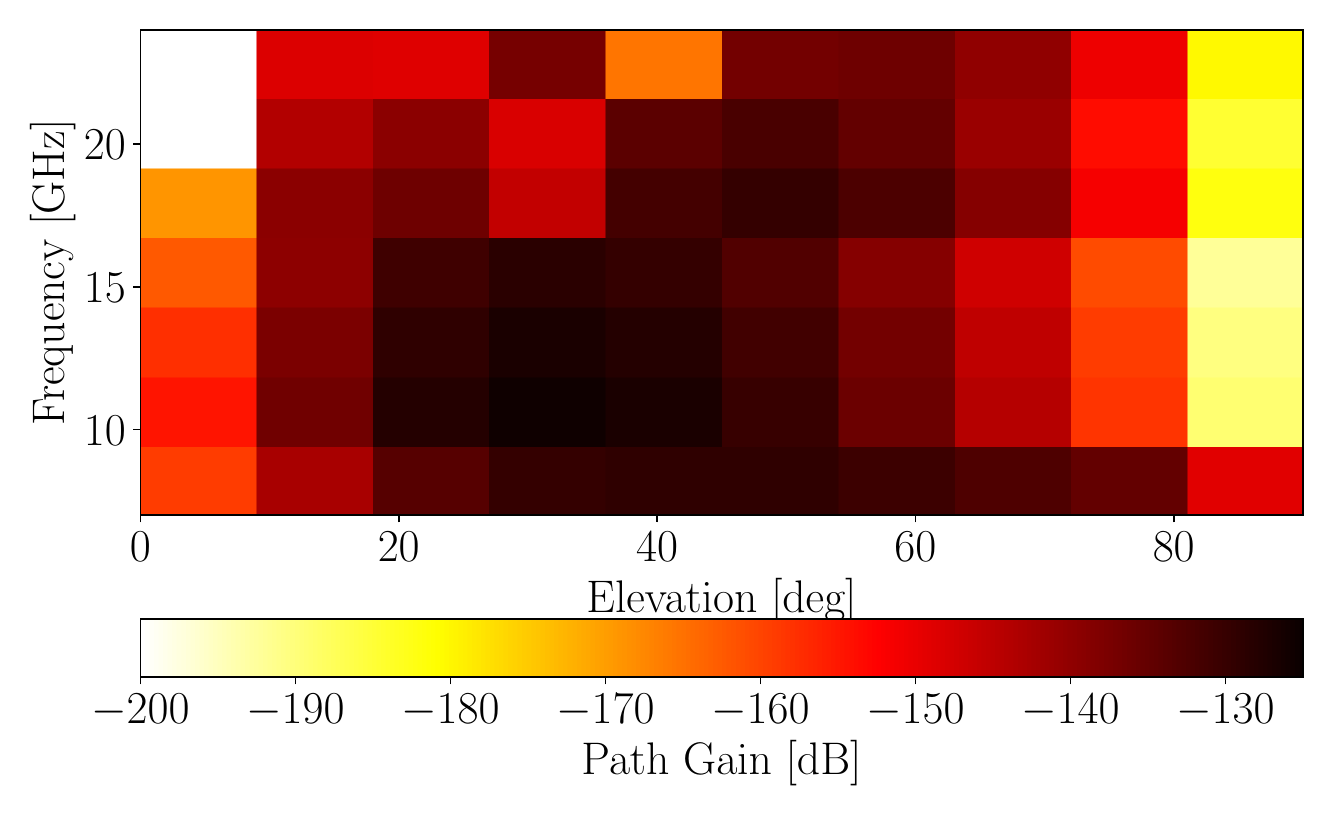}
        \caption{LoS}
        \label{fig:los}
    \end{subfigure}
    \begin{subfigure}[t]{0.65\columnwidth}
            \centering
        \includegraphics[width=\columnwidth,clip, trim=0 3.4cm 0 0]{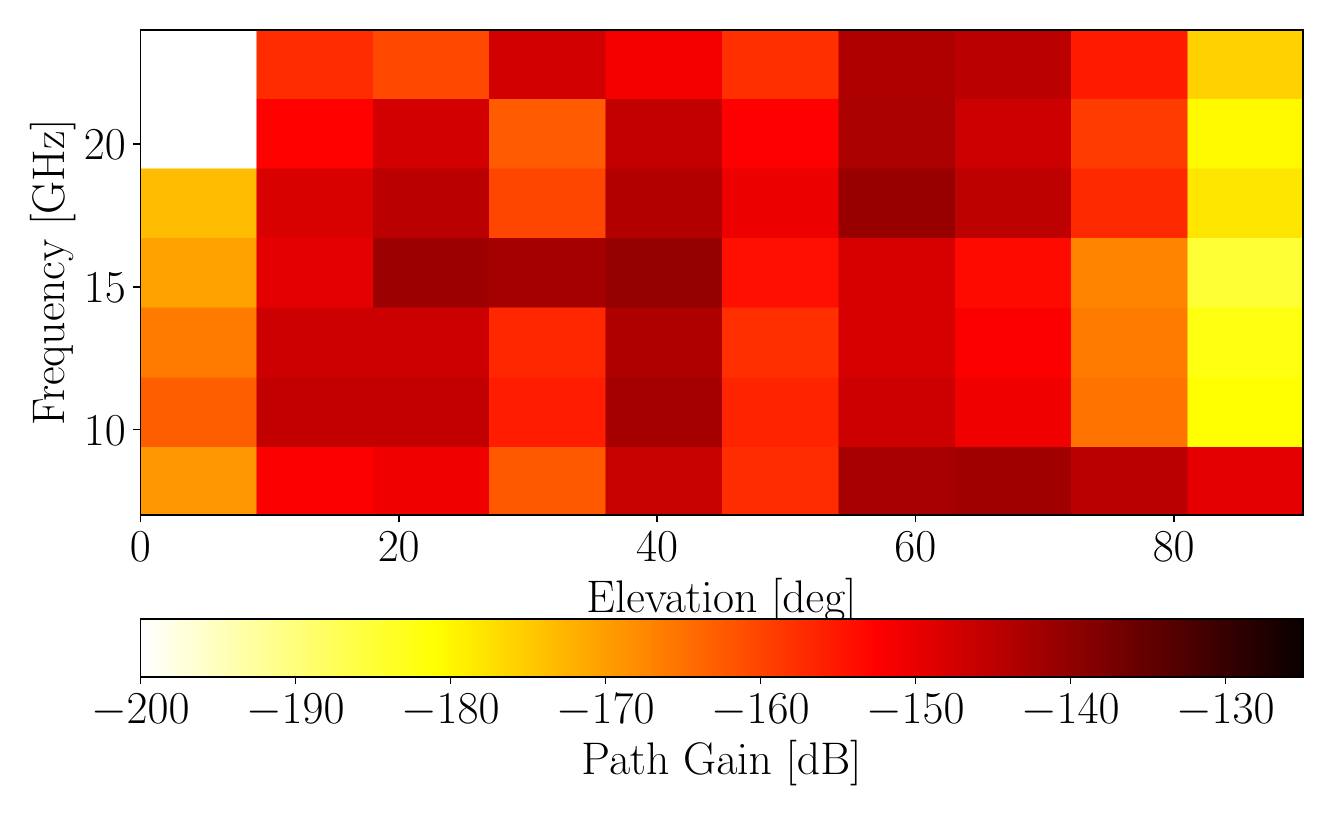}
        \caption{Reflection}
        \label{fig:refl}
    \end{subfigure}
    \begin{subfigure}[t]{0.65\columnwidth}
            \centering
        \includegraphics[width=\columnwidth,clip, trim=0 3.4cm 0 0cm]{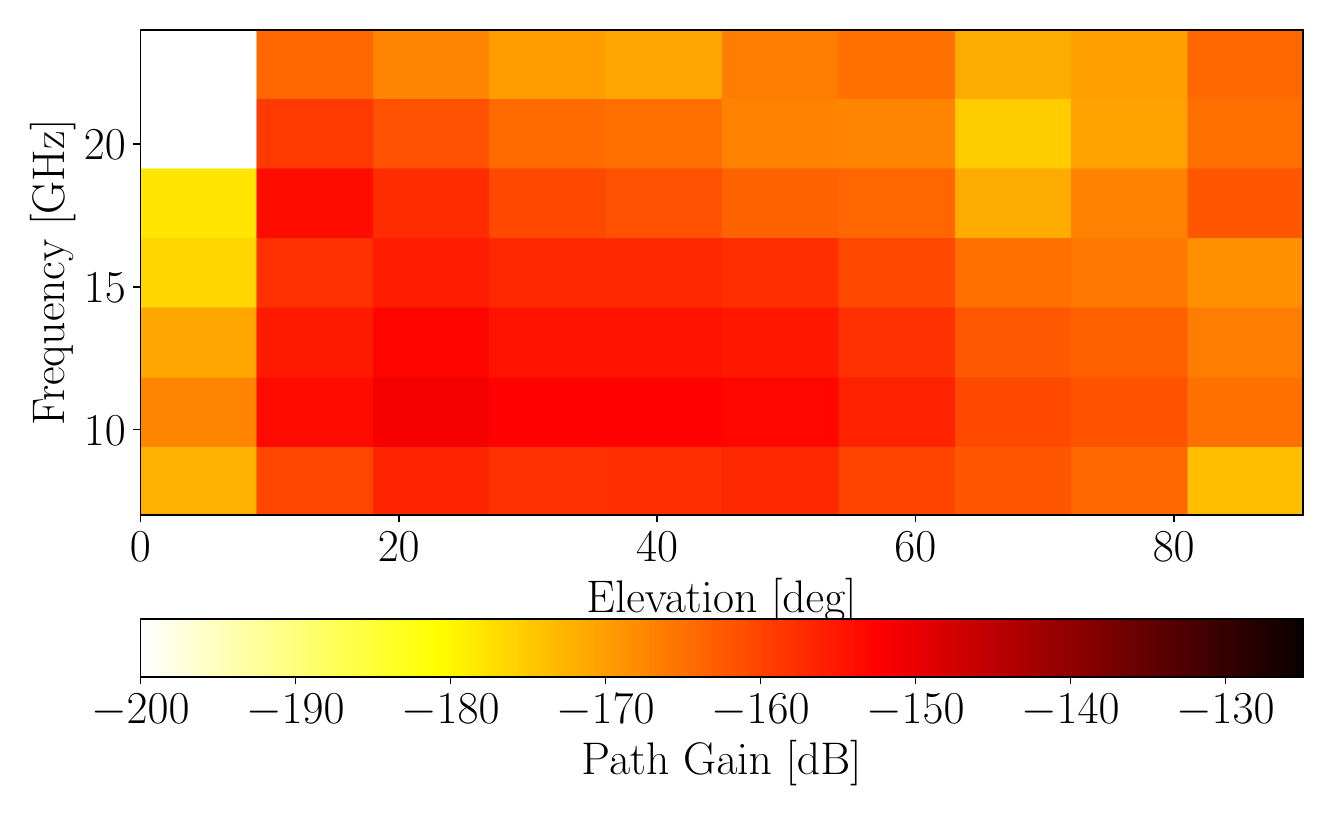}
        \caption{Diffraction}
        \label{fig:diff}
    \end{subfigure}
    \caption{Sample mean of the Path Gain plus transmitter gain $PG$ for the \gls{los} (a), reflected (b), and diffracted rays (c). As the satellite transits over the transmitters, the \gls{los} gets attenuated by the beam pattern, reproducing the trend of~\cref{fig:pattern}. Differently, the multipath components contribute the most to the aggregated interference at other satellite elevations, due to their different \gls{aod}.}
    \label{fig:ray_type}
\end{figure*}
\begin{figure*}
    \centering
    \begin{subfigure}[t]{\columnwidth}
            \centering
            \setlength\fwidth{\columnwidth}
        \setlength\fheight{.2\columnwidth}
\begin{tikzpicture}

\definecolor{red}{RGB}{255,0,0}
\definecolor{green}{RGB}{0,128,0}
\definecolor{blue}{RGB}{0,0,255}

\begin{axis}[%
 width=\fwidth,
  height=\fheight,
  hide axis,
  at={(0\fwidth,0\fheight)},
legend cell align={left},
legend style={
  fill opacity=0.8,
  draw opacity=1,
  text opacity=1,
  draw=white!15!black,
  at={(0,1)},
  anchor=north west,
  font=\footnotesize},
  xmin=10,
  xmax=50,
  ymin=0,
  ymax=0.4,
  legend columns=4,
  ]
\addlegendimage{empty legend}
\addlegendentry{Frequency: \hspace{10pt}}
\addlegendimage{red}
\addlegendentry{7 GHz}
\addlegendimage{green}
\addlegendentry{12 GHz}
\addlegendimage{blue}
\addlegendentry{24 GHz}
\end{axis}
\end{tikzpicture}
    \end{subfigure}\\
    \begin{subfigure}[t]{0.66\columnwidth}
        \centering
        \includegraphics[width=\columnwidth]{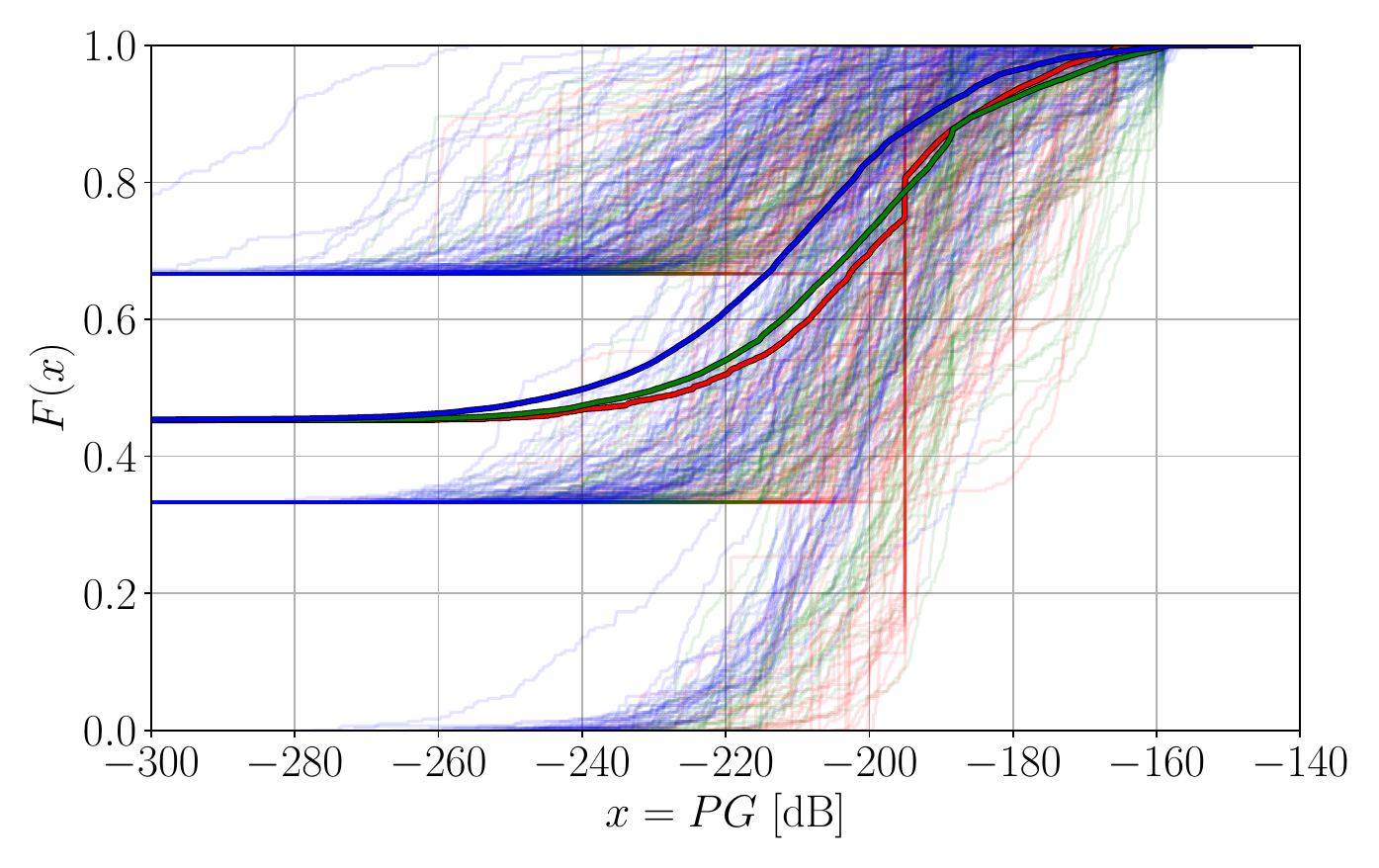}
        \caption{Satellite elevation 10\degree.}
        \label{fig:ecdf_10}
    \end{subfigure}
    \begin{subfigure}[t]{0.66\columnwidth}
        \centering
        \includegraphics[width=\columnwidth]{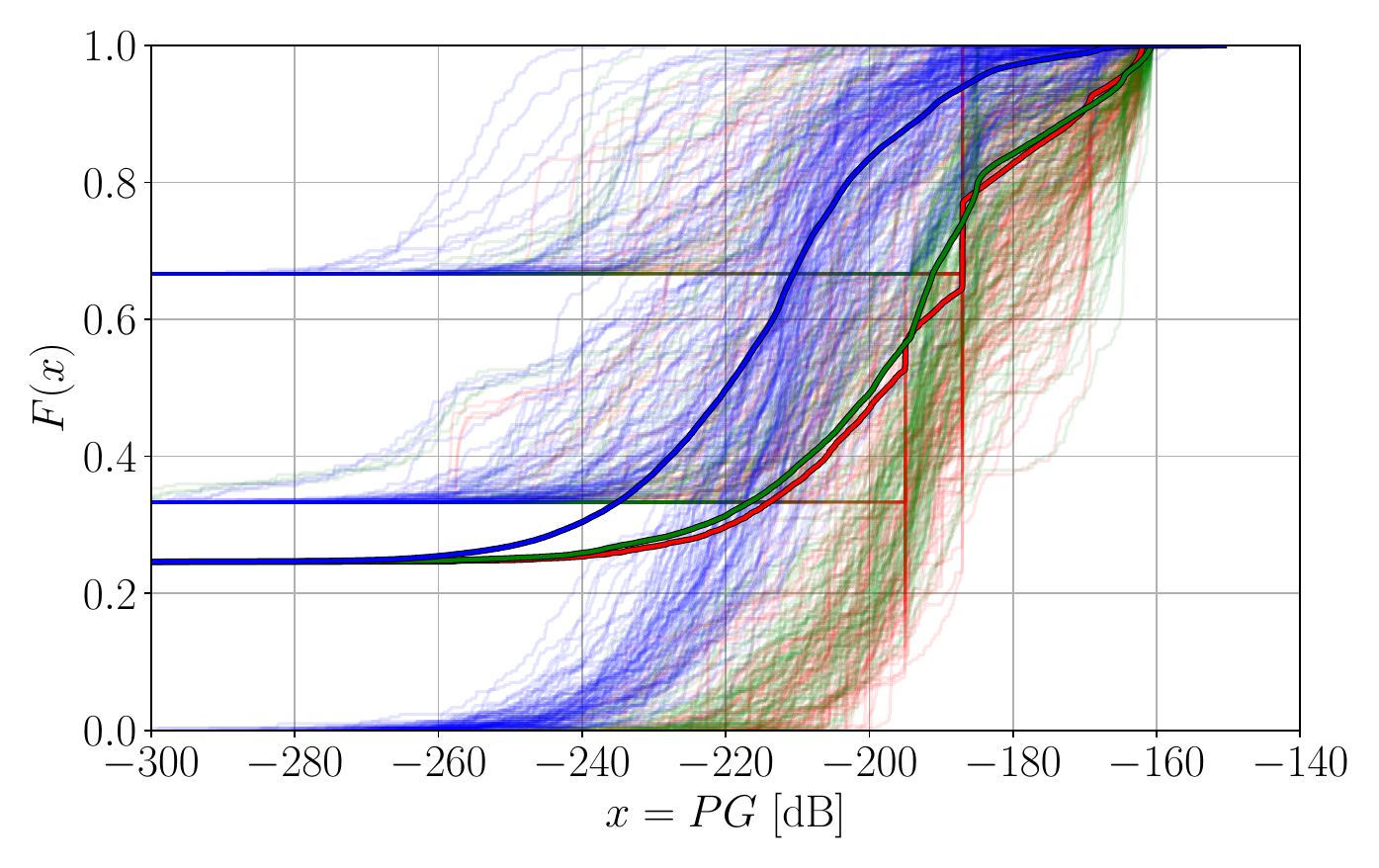}
        \caption{Satellite elevation 40\degree.}
        \label{fig:ecdf_40}
    \end{subfigure}
    \begin{subfigure}[t]{0.66\columnwidth}
        \centering
        \includegraphics[width=\columnwidth]{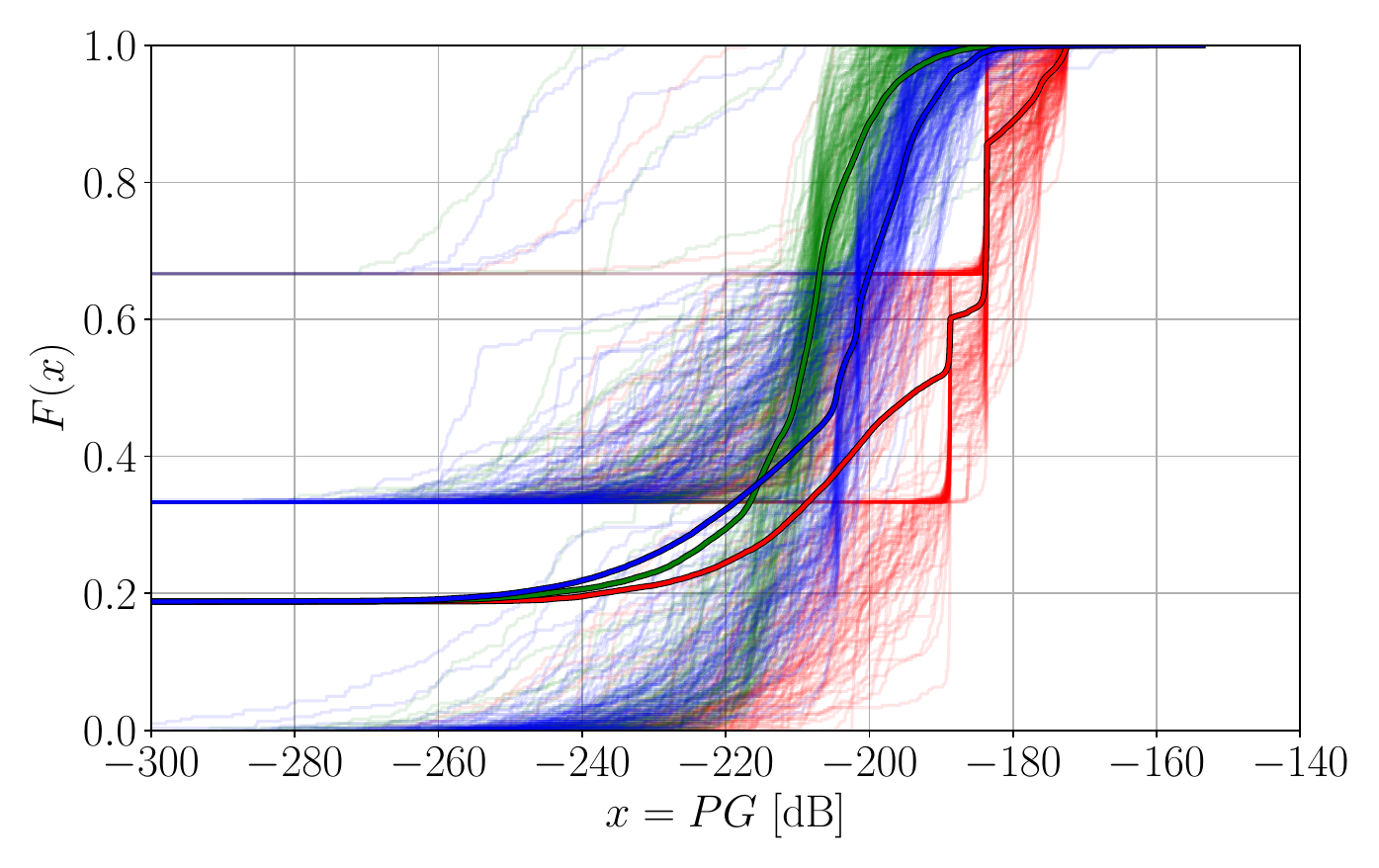}
        \caption{Satellite elevation 80\degree.}
        \label{fig:ecdf_80}
    \end{subfigure}
    \setlength\belowcaptionskip{-10pt}
    \caption{\gls{ecdf} for the considered \glspl{gnb}, obtained with 100 Monte Carlo simulations. The light lines represent the \gls{ecdf} of each \gls{gnb}, while the solid ones were obtained by accounting for all the \gls{gnb} at each frequency.}
    \label{fig:ecdf}
\end{figure*}

\section{Results}
\label{sec:results}

In this section, we first provide further details on the simulation setup and metrics of interest, and then discuss the results of the large-scale \gls{rfi} analysis.

\subsection{Simulation Setup and RFI Metrics of Interest}
The simulation parameters are reported in~\cref{tab:sim_params}.
The simulations sweep across the upper midband, focusing on the frequencies at the intersection between existing allocations and the bands of interest for the communications community, i.e., 7, 8.8, 10, 12.2, 15, 19, and 24~GHz.

A BostonTwin scene is defined by the 3D models of the structures and by the antenna locations in a given area.
For this work, we selected BOS\_G\_5, one of the default scenes, that encompasses a $1.5\times1.5$ km$^2$ area and includes 195 \glspl{gnb}.

The pattern of each antenna element follows the \gls{3gpp} standard~\cite{3gpp.38.901}.
The resulting array pattern is reported in~\cref{fig:pattern}.
As detailed in~\cref{sec:systemmodel}, at each Monte Carlo iteration, each \gls{gnb} is steered to one \gls{ue}, to simulate a realistic pointing in a dynamic system.


\begin{table}[t]
    \centering
        \caption{Simulation Parameters}
    \begin{tabular}{lll}
        \toprule
        \textbf{Parameter} & \multicolumn{2}{l}{\textbf{Value}} \\
        \midrule
        & Freq. [GHz] & \# Ant. El.\\
        \multirow{4}{*}{Array Size} & 7 & 1 \\
         & 8.8, 10, 12.2 & 4 \\
         & 15, 19 & 16 \\
         & 24 & 36 \\\hline
         Array area $A_{max}$ & \multicolumn{2}{l}{$40 \times 40$ mm$^2$} \\
         Array spacing & \multicolumn{2}{l}{$\lambda/2$}\\
         Beamforming & \multicolumn{2}{l}{Steering}\\
         Polarization & \multicolumn{2}{l}{Vertical} \\
         Element patterns & \multicolumn{2}{l}{TR 38.901\cite{3gpp.38.901}}\\\hline
         Number of transmitters & \multicolumn{2}{l}{195}\\
         Transmitter Height & \multicolumn{2}{l}{10 m}\\
         BostonTwin Scene & \multicolumn{2}{l}{BOS\_G\_5 \cite{testolina2024bostontwin}} \\
         Number of Monte Carlo iterations & \multicolumn{2}{l}{100} \\
         Minimum $PG$ & \multicolumn{2}{l}{$-120$ dB} \\
         \gls{ue}-\gls{gnb} distance & \multicolumn{2}{l}{$5-400$ m} \\\hline
         Satellite Altitude & \multicolumn{2}{l}{400 km} \\
         Satellite Elevation & 10-89\degree\\\hline
         Num. of Rays per TX & 1e7\\
         Reflection & Enabled\\
         Max. Refl. Order & 3\\
         Diffraction & Enabled\\
         \bottomrule
    \end{tabular}
    \label{tab:sim_params}
\end{table}

The broad spectrum of satellite incumbents, ranging from extremely sensitive, narrow-beam \gls{eess} radiometers to high-power radars to communications satellites, brings a correspondingly diverse set of characteristics and requirements.
Evaluating the \gls{rfi} for each incumbent requires accounting for its gain and specific requirements.
To provide more general results, we evaluate the path gain at the input of the incumbent antenna, i.e.,
\begin{equation}
    PG_{tx+ch} = \frac{E_{I}}{E_{tx}} = \sum\limits_{r=0}^{R-1}g_{tx}(\phi^D_r,\theta^D_r) a_r
    \label{eq:pg_definition},
\end{equation}
where $E_{I}$ and $E_{tx}$ are the electric field at the incumbent and at the transmitter, respectively, and $g_{tx}$ is the transmitter gain.
With this definition, we are able to study the interplay between the antenna gain and directionality with the channel and the relevant propagation phenomena.
Note that commonly, satellites employ directional antennas to compensate for the increased path loss (in communications) and to reduce the observation area and increase the spatial resolution (both communications and scientific purposes).
However, given the satellite's altitude, even a narrow beam projects a large footprint on the ground.
Note that any emission from the footprint area is amplified by the receiver antenna within $-3$~dB from the peak gain $G_{I} = \max\{g_I(\phi,\theta)\}$, independently of the \gls{aoa}.
Thus, extending the results presented in this work to a specific incumbent with gain $G_I$ is straightforward, or considering a transmit power $P_{tx}$, as the interference power $P_I$ is obtained as 
\begin{equation}
    P_I = PG_{tx+ch} G_{I} P_{tx}.
\end{equation}
In the following, we drop the subscript from $PG_{tx+ch}$ for readability, and we use $PG$ to refer to the quantity defined in \cref{eq:pg_definition}.

\subsection{Relevant Propagation Phenomena in a Ground-to-Space Channel Model}
\label{ssec:ray_type}


We first analyze the contribution of the three propagation modes, i.e., \gls{los}, reflection, and diffraction, to the aggregated interference.
For tractability, diffuse scattering is not considered in this work.

Figure~\ref{fig:ray_type} shows the sample mean of the path gain $PG_{tx+ch}$ for different midband frequencies and satellite elevation angles, for the \gls{los} ($PG_{LoS}$), the reflected ($PG_{refl}$), and the diffracted rays ($PG_{diff}$), computed as follows:
\begin{align}
    PG_{type} &= \sum\limits_{b=0}^{N_{gNB}-1}PG_b = \sum\limits_{b=0}^{N_{gNB}-1}\sum\limits_{r=0}^{R_b-1}g_{b}(\phi^D_r,\theta^D_r) a_r\delta_{type},
\end{align}
where \( R_b \) is the number of rays (propagation paths) from the \( b^{th} \) \gls{gnb} to the satellite, \( g_b(\phi_r^D, \theta_r^D) \) is the \gls{gnb} gain in the direction \( (\phi_r^D, \theta_r^D) \) of the \( r ^{th}\) ray.

\( a_r \) is the attenuation factor of the \( r \)th ray due to propagation effects, as defined in~\cref{eq:total_abs}, and
\begin{equation}
    \delta_{type} = \begin{cases}
        1 & \text{if ray type is $type$}\\
        0 & \text{otherwise}
    \end{cases}, type\in\{\text{LoS, refl., diff.}\}.
\end{equation}

The bottom row of~\cref{fig:ray_type} (a-c) represents the path gain at 7~GHz, obtained using a single patch antenna with low directionality~\cite{3gpp.38.901}, as reported in~\cref{fig:pattern}.
Thus, it can be used as a baseline to understand the directionality of the channel.
As the satellite rises over the horizon and its elevation angle increases, the distance between the ground transmitters and the incumbent decreases.
Thus, the free-space and the atmospheric loss decrease, as the signal travels a shorter distance and crosses fewer layers of the atmosphere.
At the same time, when the satellite is low on the horizon, the probability of blockage by buildings is higher than when the satellite is above the city.
As the incumbent transits over the city, the interference increases, peaking around $40\degree$ elevation for the \gls{los} channel, when more interferers are in \gls{los} and before getting attenuated by the antenna pattern.
The reflected and diffracted rays in \cref{fig:refl} and~\ref{fig:diff} exhibit similar behavior.

When using directional arrays, steering the \gls{gnb} arrays toward the users amplifies the \gls{los} path for lower elevation angles and suppresses it when the satellite is above the network.
This is visible in the last column of \cref{fig:los}, where the aggregated path gain does not cross the $-180$~dB threshold.
Further increasing the directionality, e.g., using 16 (15, 19~GHz) or 36 (24~GHz) antenna elements, effectively reduces the interference peak, the \gls{los} toward a satellite at $30-40\degree$ elevation is attenuated by the null of the array pattern (\cref{fig:pattern}).
However, the presence of sidelobes limits the effectiveness of the directionality, as the interference around $50\degree$ elevation remains significant.

The reflected rays (\cref{fig:refl}) exhibit a distinct trend with peaks around $40\degree$ elevation, strongly linked to geometric conditions.
Specifically, signals directed downward toward terrestrial users are reflected from scatterers, e.g., buildings and roads, aligning optimally at mid-elevation angles.

At low to mid-elevation angles ($10^\degree$–$60^\degree$) and lower frequencies, the signal is diffracted by the building edges in its path, reaching the incumbent with non-negligible power.

Moreover, in general, the \gls{aod} of the multipath components differs from that of the \gls{los}.
Correspondingly, the transmit gain applied to the reflected rays is different than that of the \gls{los}, which results in the different patterns visible in \cref{fig:refl} and \cref{fig:diff}.
Notably, more interference is generated by the multipath components than the \gls{los} when the satellite is above the terrestrial network ($89\degree$ elevation), particularly for highly directional arrays, as they can be amplified by the main lobe before bouncing towards the satellite.
This is in line with our previous findings~\cite{testolina2024modeling}, which showed that \gls{rfi} analysis for very directional networks should take into account secondary paths to the incumbent.

\subsection{Interference by a single \gls{gnb}}
\label{ssec:single_gnb}
Figure~\ref{fig:ecdf} shows the \gls{ecdf} of the path gain for each \gls{gnb} $PG_b$, $b\in\left\{0,..,N_{gNB}\right\}$ obtained with 100 Monte Carlo simulations at 7 (red), 12 (green), and 24 GHz (blue) for three satellite elevation angles.
The semi-transparent lines represent the individual \gls{gnb} distributions, while the opaque ones represent the aggregated statistics, i.e., the distributions of all the collected samples at the specified frequency and elevation angle.
Note that, at each Monte Carlo iteration, the \glspl{gnb} are steered toward a random user among the selected ones, as explained in~\cref{sec:systemmodel}.

At low satellite elevation angles, such as in ~\cref{fig:ecdf_10}, a significant overlap among different frequencies and \glspl{gnb} is exhibited, suggesting that interference at lower angles is primarily determined by blockage and propagation distance rather than frequency-dependent antenna beamforming. Nevertheless, the probability of the \gls{gnb} being steered toward the incumbent is greater than at larger elevation angles. For this reason, the rightmost tails of the 24~GHz distributions emerge from the others, corresponding to the cases when beamforming amplifies the interference in the direction of the satellite.

With a single antenna (7~GHz, red line), the \gls{gnb} \glspl{ecdf} are often step-shaped, indicating a low variability across the Monte Carlo iterations.
A low-directivity antenna minimizes variation across iterations, as the primary difference between iterations lies in the direction of the \glspl{gnb} antenna.
Similarly, as the satellite elevation increases, the beamforming used at 12 and 24~GHz effectively suppresses the \gls{rfi} to the satellite, reducing the breadth of the corresponding \glspl{ecdf}.

At moderate elevations, such as in~\cref{fig:ecdf_40}, the 12 GHz configuration generates higher interference than the 24 GHz configuration, due to the lower elevation of the sidelobes of the $4\times4$ (12 GHz) than those of the $6\times6$ (24 GHz) array.
Conversely, at high elevations (\cref{fig:ecdf_80}) the interference at 24 GHz exceeds that at 12 GHz.

All the distributions show a plateau lower tail, corresponding to the samples where the generated interference is below the numerical precision of the simulation, i.e., $PG\simeq-\infty$.
The plateau value of $F(x)$ represents the corresponding probability for the single \gls{gnb} (transparent) and on the whole network (opaque).
For the former, there appear to be three clusters of \glspl{gnb}, characterized by the zero-interference probability of 0.67, 0.38, and 0.
As the satellite rises over the horizon, the probability of generating non-negligible interference increases, and thus the \glspl{gnb} move from the lower to the higher-probability clusters.
Similarly, the overall probability of causing zero interference decreases from 0.43 (10\degree) to 0.23 (40\degree) and 0.2 (80\degree).
This suggests that there exists \glspl{gnb} that due to their placement are more prone to generating significant \gls{rfi}, e.g., those in the first cluster.
Site-specific sharing systems can take this feature into account to minimize changes to the ground network and reduce the coordination required to meet the safety interference margins.

\begin{figure}
    \centering
    \setlength\fheight{0.5\columnwidth}
    \setlength\fwidth{\columnwidth}
    \includegraphics[width=\fwidth,clip, trim=0 0 0 2cm]{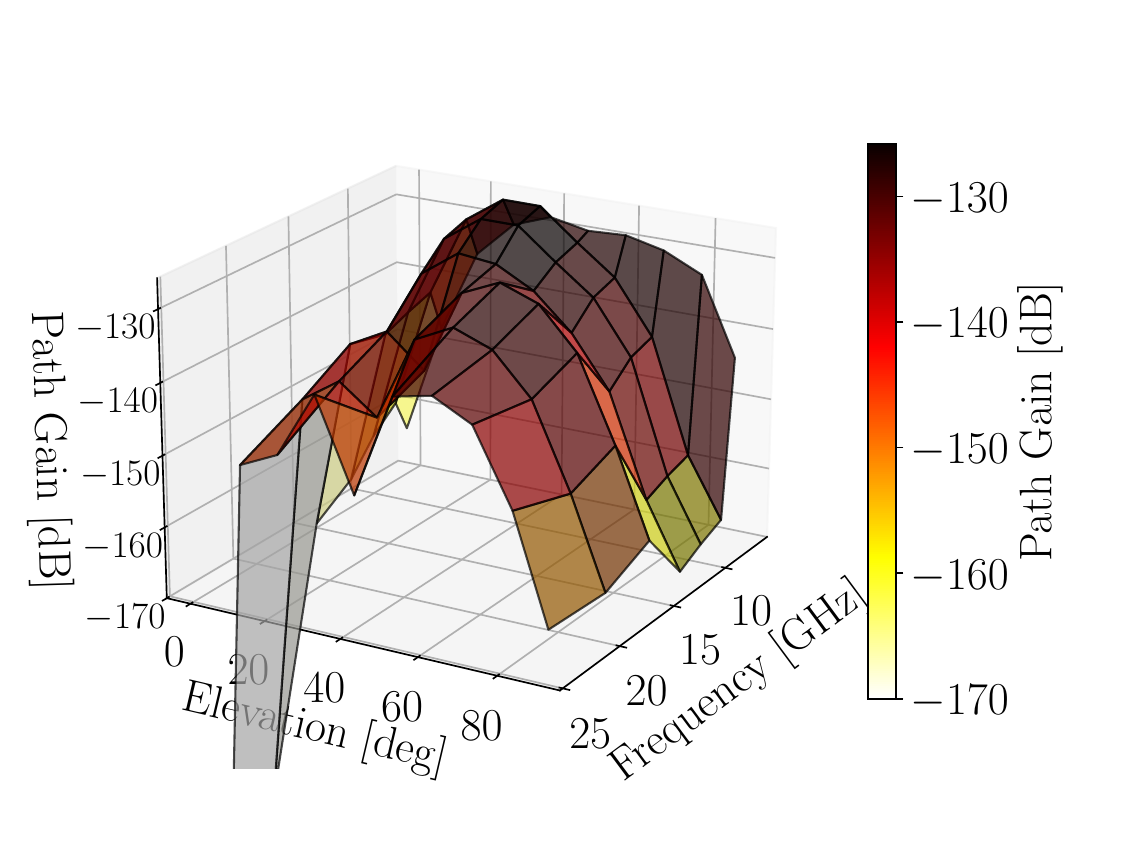}
    \caption{Aggregated path gain at the satellite incumbent. Directional beams can effectively suppress the interference when the incumbent transits above the transmitter. When the satellite is high over the horizon (30-50\degree), the sidelobes of the array cause a significant leakage toward the incumbent.}
    \label{fig:aggregated_pg_3d}
\end{figure}

\subsection{Aggregated Interference}
Figure~\ref{fig:aggregated_pg_3d} reports the aggregated path gain from the \glspl{gnb} in the considered area, averaged over the 100 Monte Carlo iterations and for different frequencies and satellite elevation angles.
The surface mirrors the results presented in~\cref{ssec:ray_type} and~\cref{ssec:single_gnb}.
At $0\degree$, the atmosphere and the satellite distance effectively reduce \gls{rfi}, even when aggregated over multiple sources. The transmitter beamforming plays a similar role for $90\degree$ elevations.
However, if the transmitter pattern is not directional enough ($7-8$~GHz), or has significant sidelobes ($15-24$~GHz), the transmit power can leak in the direction of the incumbent, peaking at around $-125$~dB when using 4 antennas at $8.8-12$~GHz.
We remark that the chosen metric does not take into account the transmit power of the transmitters, nor the incumbent gain.
Considering the high directivity of the antennas onboard satellites, and the extreme sensitivity of the scientific radiometers used by \gls{eess}, the safety levels defined by the \gls{itu} recommendations could be easily exceeded.
However, narrower beams show promising results, suggesting that a careful beam design, which suppresses the rays in the direction of the incumbent, can enable the sharing of the upper mid-band between future terrestrial networks with existing satellite services. 



\section{Conclusions and Future Work}
The wireless community is 
analyzing the risks and opportunities associated with spectrum sharing and coexistence with existing services in the upper midband, a candidate for next-generation 6G systems.
In this work, we considered a realistic deployment of a terrestrial network working at upper midband frequencies ($7-24$~GHz), along with a generic satellite incumbent.
Using the Sionna \gls{rt} and BostonTwin, we modeled the reflections and diffractions of the transmitted signals on the 3D models of the buildings in the considered $1.5\times 1.5$ km area in Boston, USA.
Additionally, the effect of atmospheric absorption were introduced.
We analyzed separately the contribution of the multipath components generated by the interaction with the environment elements, and the interplay with directional antennas.
The results showed that the power reaching a \gls{leo} incumbent from a ground network and a realistic deployment can sum up to a significant level of interference.
Directional antennas and beamforming are shown to mitigate the \gls{rfi} for some satellite elevation angles, but the power leaking through the sidelobes remains significant.
Therefore, we concluded that directionality alone was insufficient to meet international interference requirements, highlighting the need for carefully designed beam patterns and coordinated deployment strategies.

Furthermore, the scalability of our framework, enabled by Sionna, allows for large-scale evaluation of coexistence scenarios. This feature is crucial for assessing the feasibility of 6G deployments across diverse urban environments and ensuring that findings generalize to realistic network conditions. 

As future work, we plan to further improve the channel model, by accounting for other elements such as foliage loss and scattering, as well as different satellite orbits and terrestrial network configurations.
Furthermore, the obtained results indicate the possibility of designing terrestrial networks strategically to minimize interference with existing satellite services, paving the way to a safe, coexistence-oriented \gls{6g} deployment.

\bibliographystyle{IEEEtran}
\bibliography{biblio}

\end{document}